\documentclass[twocolumn,preprintnumbers,superscriptaddress,endnote,nofootinbib,aps,prd,floatfix]{revtex4}

\usepackage{graphicx,amsmath,amssymb}
\usepackage{slashed,subfigure}
\usepackage{bbm}
\usepackage{footmisc}
\usepackage{booktabs}
\usepackage{ulem}

\usepackage{hyperref}
\hypersetup{colorlinks=true, citecolor=blue, urlcolor=blue, linkcolor=blue}

\newcommand{\be}{\begin{eqnarray*}}
\newcommand{\ee}{\end{eqnarray*}}

\newcommand{\bee}{\begin{eqnarray}}
\newcommand{\eee}{\end{eqnarray}}
\newcommand{\beeq}{\begin{equation}}
\newcommand{\eeeq}{\end{equation}}

\newcommand{\beq}{\begin{eqnarray}} 
\newcommand{\eeq}{\end{eqnarray}} 
\newcommand{\non}{\nonumber} 
\newcommand{\MSb}{\overline{\text{MS}}}

\begin{document}

\title{Di-Higgs Peaks and Top Valleys: Interference Effects in Higgs Sector Extensions}

\begin{abstract}
In models with extended scalars and CP violation, resonance searches in double Higgs final states stand in competition with related searches in top quark final states as optimal channels for the discovery of beyond the Standard Model (BSM) physics. This complementarity is particularly relevant for benchmark scenarios that aim to highlight multi-Higgs production as a standard candle 
for the study of BSM phenomena. In this note, we compare interference
effects in  $t\bar t$ final states with correlated phenomena in double
Higgs production in the complex  singlet and the complex
2-Higgs-Doublet Models. Our results indicate that the  BSM discovery
potential in  di-Higgs searches can be underestimated in comparison to
$t\bar t $ resonance searches. Top pair final states are typically
suppressed due to destructive signal-background interference, while
$hh$ final states can be enhanced due to signal-signal
interference. For parameter choices where the two heavy Higgs
resonances are well separated in mass, top final states are
suppressed relative to the naive signal expectation,
while estimates of the production cross section times branching ratio remain accurate at the
${\cal{O}}(10\%)$ level for double Higgs final states. 
\end{abstract}

\author{Philipp Basler}\email{philipp.basler@kit.edu}
\affiliation{Institute for Theoretical Physics, Karlsruhe Institute of Technology, 76128 Karlsruhe, Germany.\\[0.1cm]}

\author{Sally Dawson}\email{dawson@bnl.gov}
\affiliation{Department of Physics, Brookhaven National Laboratory, Upton, N.Y., 11973, U.S.A.\\[0.1cm]}

\author{Christoph Englert}\email{christoph.englert@glasgow.ac.uk}
\affiliation{SUPA, School of Physics and Astronomy, University of Glasgow, Glasgow G12 8QQ, U.K.\\[0.1cm]}

\author{Margarete
  M\"uhlleitner}\email{milada.muehlleitner@kit.edu}
\affiliation{Institute for Theoretical Physics, Karlsruhe Institute of Technology, 76128 Karlsruhe, Germany.\\[0.1cm]}
\vspace*{0.1cm}

\maketitle

\section{Introduction}
\label{sec:intro}
The search for new physics beyond the Standard Model remains a priority of the Large Hadron Collider (LHC) phenomenology programme. Although model-independent search strategies are gaining momentum, concrete and well-motivated UV scenarios still provide vital information on why new physics has not been observed so far. Recent investigations~\cite{Bahl:2018zmf,Basler:2018dac,Baum:2019pqc,Babu:2018uik} highlight the point that the non-observation of new physics can be re\-con\-ciled with ``standard'' Higgs sector extensions far away from their decoupling limits. This is possible when the non-Standard Model (non-SM) degrees of freedom have suppressed production cross sections, or when they are hidden in channels that are experimentally difficult to observe. Particularly relevant in this context are di-Higgs final states that might act as the main discovery channel  for BSM physics in such an instance.

The largest production channel for electroweak scalars that have significant top couplings proceeds through gluon fusion at the LHC, see {\it e.g.}~\cite{Dittmaier:2011ti}. This directly motivates resonance searches in top final states if they are kinematically accessible. It is known that resonance searches in top pairs are particularly vulnerable to large interference effects~\cite{Gaemers:1984sj,Dicus:1994bm,Bernreuther:1998qv,Jung:2015gta,Frederix:2007gi,Barger:2006hm,Craig:2015jba,Bernreuther:2015fts,Carena:2016npr,Hespel:2016qaf,BuarqueFranzosi:2017qlm,BuarqueFranzosi:2017jrj,Djouadi:2019cbm,Kauer:2019qei} that can have a significant impact on the formulation of exclusion constraints~\cite{Aaboud:2017hnm,Brooijmans:2018xbu,Sirunyan:2019wph,DiMicco:2019ngk}. It is therefore entirely possible that a new (possibly gauge-phobic) scalar is not visible as  an isolated $t\bar t$ resonance. Under these circumstances, multi-Higgs production becomes a crucial tool for BSM discovery.

It is the purpose of this paper to quantitatively compare $t\bar t$
and $hh$ resonance searches for models where we would expect top
production to play the leading role in a new physics discovery. This
is particularly highlighted in  the complex singlet (CxSM)
and CP-violating  2-Higgs-Doublet Models (C2HDM). Building on previous
insights~\cite{Basler:2018dac}, we show that in the relevant regions
of the C2HDM 
interference effects can lead to misleading sensitivity estimates. In particular, the sensitivity from di-Higgs final states~\cite{Dawson:2015haa}
can be underestimated relative to top pair production, {\it{i.e.}} destructive interference
effects in $t\bar t$ searches correlate with constructive interference in 
$hh$ searches in parameter regions of the C2HDM that are allowed in the light of current LHC searches. We contrast these findings against results obtained for the CxSM.

This work is organised as follows: In Sec.~\ref{sec:models}, we
briefly review the CxSM and the C2HDM that we consider in this work. We provide details of our calculation with
results in Sec.~\ref{sec:res}, where we also present a number of benchmark
points that highlight the phenomenology. We conclude in Sec.~\ref{sec:conc}. 

\section{Models and Scans}
\label{sec:models}
\subsection{The CxSM}
\label{sec:cxsm}
The CxSM is based upon the extension of the SM by a complex scalar
field \cite{Barger:2008jx,Gonderinger:2012rd,Coimbra:2013qq,Jiang:2015cwa,Sannino:2015wka,Costa:2015llh,Muhlleitner:2017dkd,Chiang:2017nmu,Azevedo:2018llq}. The CxSM potential with a softly broken global $U(1)$
symmetry is given by,
\beq
V &=& \frac{m^2}{2} H^\dagger H + \frac{\lambda}{4} (H^\dagger
H)^2+\frac{\delta_2}{2} H^\dagger H |\mathbb{S}|^2 +
\frac{b_2}{2}|\mathbb{S}|^2 \nonumber \\
&+& \frac{d_2}{4} |\mathbb{S}|^4 +
\left(\frac{b_1}{4} \mathbb{S}^2 + a_1 \mathbb{S} +c.c. \right)
\, ,  \label{eq:VCxSM}
\eeq
with 
\beq
\mathbb{S} =  S + iA
\eeq 
being a hypercharge zero scalar field and the soft breaking
terms being written in parenthesis. After electroweak symmetry
breaking the fields can be written as  
\begin{equation}
H=\dfrac{1}{\sqrt{2}}\left(\begin{array}{c} G^+ \\
    v+h+iG^0\end{array}\right) 
\end{equation}
and
\begin{equation}
\mathbb{S}=\dfrac{1}{\sqrt{2}}\left[v_S+s+i(v_A+ a)\right] \;,
\label{eq:fieldsCxSM}
\end{equation}
where $v\approx 246$~GeV is the SM vacuum expectation value (VEV) of
the $h$ field and $v_S$ and $v_A$ are the VEVs of the real and
imaginary parts of the complex singlet field, respectively. The
hermiticity of the potential implies that all parameters are real, 
except for the soft breaking terms. We also impose invariance under 
$\mathbb{S} \to \mathbb{S}^*$ (or $A \to -A$) so that  $a_1$ and $b_1$ are real.
This implies that the theory is described by $7$ independent parameters.
The model can be studied by treating the real and 
imaginary components of the complex singlet as independent fields, which implies that the model
is equivalent to one with $2$ real singlets and has no CP violation.
For our investigation we choose to work  in the broken phase where all three
VEVs are non-zero, as this phase implies mixing between all three
CP-even scalars. Their mass eigenstates $H_i$ ($i=1,2,3$) are obtained
from the gauge eigenstates through the rotation matrix $R$ parametrised as
\beq
R =\left( \begin{array}{ccc}
c_{1} c_{2} & s_{1} c_{2} & s_{2}\\
-(c_{1} s_{2} s_{3} + s_{1} c_{3})
& c_{1} c_{3} - s_{1} s_{2} s_{3}
& c_{2} s_{3} \\
- c_{1} s_{2} c_{3} + s_{1} s_{3} &
-(c_{1} s_{3} + s_{1} s_{2} c_{3})
& c_{2}  c_{3}
\end{array} \right) \;,
\label{eq:rotsinglet}
\eeq
where we have introduced the shorthand notation $s_{i} \equiv \sin
\alpha_i$ and $c_{i} \equiv \cos \alpha_i$. Without loss of generality
we vary the angles in the range
\beq
-\frac{\pi}{2} \le \alpha_i < \frac{\pi}{2} \,.
\label{eg:alpharanges1}
\eeq
The masses of the neutral Higgs bosons are ordered as $m_{H_1} \leq
m_{H_2} \leq m_{H_3}$. As input parameters we choose the set
\beq
\alpha_1 \;, \quad \alpha_2 \;, \quad \alpha_3 \;, \quad v \;, \quad v_S \;,
\quad m_{H_1} \quad \mbox{and} \quad m_{H_3} \;. \label{eq:cxsminput}
\eeq
The remaining parameters are determined internally in
{\tt ScannerS} \cite{Coimbra:2013qq,Ferreira:2014dya}, with which we perform
our scan in the parameter space of the model,  taking into account
the minimum conditions on the vacuum. 

\subsection{The CxSM Scan}
\label{sec:cxsmscan}
In order to find viable points in the parameter space of the CxSM that
are compatible with the relevant theoretical and experimental
constraints, we performed a scan using the program {\tt
  ScannerS}. The program checks for  theoretical constraints such
as the requirement of the potential to be bounded from below, the
chosen vacuum to be a global minimum and perturbative unitarity to be
fulfilled. We furthermore require that the mass of
the lightest Higgs boson, identified with the SM-like one and
denoted by $h$, is $m_h=125.09$~GeV~\cite{Aad:2015zhl}. 
Compatibility with the electroweak precision data is ensured by
applying a 95\% C.L.~exclusion limit from the electroweak precision
observables $S$, $T$ and $U$ \cite{Peskin:1991sw,Maksymyk:1993zm}, see
\cite{Costa:2014qga} for further details. Compatibility with the
exclusion limits from the collider data on Higgs observables at 95\%
C.L.~has been checked by using {\tt
  HiggsBounds5.2.0}~\cite{Bechtle:2008jh,Bechtle:2011sb,Bechtle:2013wla} and compatibility
  with the Higgs rates was verified by using {\tt
    HiggsSignals2.2.1}~\cite{Bechtle:2013xfa}. 
The necessary production cross sections were obtained from {\tt
    ScannerS} which uses results from {\tt SusHi1.6.1}
  \cite{Harlander:2012pb,Harlander:2016hcx}.
 The required branching ratios to compute
the signal strength were computed with {\tt
  sHDECAY} \cite{Costa:2015llh} which is based on the implementation of
the CxSM and also the real singlet extension of the SM (RxSM) both in
their symmetric and broken phases in 
{\tt HDECAY}~\cite{Djouadi:1997yw,Djouadi:2018xqq}. 

The SM input
parameters are chosen as~\cite{PhysRevD.98.030001,Dennerlhcnote} 
\begin{equation}
\begin{tabular}{lcllcl}
\quad $\alpha(M_Z)$ &=& 1/127.92, & $\alpha^{\MSb}_s(M_Z)$ &=&
0.118, \\
\quad $M_Z$ &=& 91.187~GeV, & $M_W$ &=& 80.358~GeV,  \\
\quad $m_t$ &=& 172.5~GeV, & $m^{\MSb}_b(m_b^{\MSb})$ &=& 4.18~GeV, \\
\quad $m_\tau$ &=& 1.777~GeV. 
\end{tabular}
\end{equation}
The remaining light quark  and lepton masses have been set to
\cite{PhysRevD.98.030001,Dennerlhcnote} 
\beq
\label{eq:inputend}
\begin{array}{lcllcl}
m_e &=& 0.5110 \mbox{ MeV} \;, & m_\mu &=& 105.66  
\mbox{ MeV} \;, \\ 
m_u &=& 100 \mbox{ MeV} \;, & m_d &=& 100 \mbox{ MeV} \;, \\
m_s &=& 100 \mbox{ MeV} \;.
\end{array}
\eeq

Our sample points were generated with the input parameters listed in
Eqs.~(\ref{eq:cxsminput})-(\ref{eq:inputend}). Identifying the lightest
  Higgs boson with the SM-like Higgs boson $h$, the remaining ones
are restricted to the mass range 
\beq
{125.09} \; \mbox{GeV } {<} m_{H_i} {\le} 1000 \;\mbox{GeV} ,\;H_i \ne h \;.
\eeq
The VEVs $v_A$ and $v_S$ are varied in the range\footnote{Note that $v_A$ is not varied as independent input parameter, we solely make sure that by our choice of input parameters $v_A$ lies in the range defined in Eq.~\eqref{eq:vev}.}
\beq
\label{eq:vev}
1 \mbox{ GeV } \le v_A, v_S < {1.0} \mbox{ TeV} \;,
\eeq
and the  mixing angles as in Eq.~(\ref{eg:alpharanges1}).
All input parameters except for the mixing angles were 
generated randomly and uniformly in the ranges
specified above. The mixing angles 
on the other hand were extracted from the mixing matrix elements
of $R$ defined in Eq.~(\ref{eq:rotsinglet}), that were generated uniformly. 
Through this procedure the couplings to fermions and gauge bosons are
distributed uniformly. 

We also checked if the parameters of the final data set induce a strong
first order phase transition, which is a necessary condition for successful 
baryogenesis~\cite{Sakharov:1967dj,Quiros:1994dr,Moore:1998swa}, by
using the {\tt C++} code {\tt BSMPT}~\cite{Basler:2018cwe}. 
We found that none of the benchmark points satisfies a
  strong first order phase transition.
 
\subsection{The C2HDM}
By adding a second $SU(2)_L$ Higgs doublet to the SM Higgs sector we
obtain the 2-Higgs-Doublet Model (2HDM)
\cite{Lee:1973iz,Gunion:1989we,Branco:2011iw}. Imposing a
$\mathbb{Z}_2$ symmetry, under which the two $SU(2)_L$ doublets
$\Phi_j$ ($j=1,2$) transform as $\Phi_1 \to \Phi_1$ and $\Phi_2 \to
-\Phi_2$, the Higgs potential of a general 2HDM with the $\mathbb{Z}_2$
symmetry softly broken can be cast into the form
\beq
V &=& m_{11}^2 |\Phi_1|^2 + m_{22}^2 |\Phi_2|^2 \non\\
&-& (m_{12}^2 \Phi_1^\dagger
\Phi_2 + h.c.) + \frac{\lambda_1}{2} (\Phi_1^\dagger \Phi_1)^2 \non\\
&+& \frac{\lambda_2}{2} (\Phi_2^\dagger \Phi_2)^2 
+ \lambda_3 (\Phi_1^\dagger \Phi_1) (\Phi_2^\dagger \Phi_2) \non\\
&+&  \lambda_4 (\Phi_1^\dagger \Phi_2) (\Phi_2^\dagger \Phi_1) +
\left[\frac{\lambda_5}{2} (\Phi_1^\dagger \Phi_2)^2 + h.c.\right] \; .
\eeq
The $\mathbb{Z}_2$ symmetry is extended to the fermion sector, thereby
guaranteeing the absence of flavour-changing neutral currents. The
$\mathbb{Z}_2$ charge assignments can be distributed such that we
obtain four phenomenologically different 2HDM types 
summarized in Tab.~\ref{tab:four2hdmtypes}. 
\begin{table}
\begin{center}
\begin{tabular}{r|ccc} \toprule
& $u$-type & $d$-type & leptons \\ \hline
type I (T1) & $\Phi_2$ & $\Phi_2$ & $\Phi_2$ \\
type II (T2) & $\Phi_2$ & $\Phi_1$ & $\Phi_1$ \\
lepton-specific & $\Phi_2$ & $\Phi_2$ & $\Phi_1$ \\
flipped & $\Phi_2$ & $\Phi_1$ & $\Phi_2$ \\ \bottomrule
\end{tabular}
\caption{The four Yukawa types of the softly broken
$\mathbb{Z}_2$-symmetric 2HDM, 
  defined by the Higgs doublet that couples to each kind of
  fermions. \label{tab:four2hdmtypes}} 
\end{center}
\end{table}
For the Higgs potential to be Hermitian, all parameters must be real,
with exception of $\lambda_5$ and $m_{12}^2$. We obtain the complex or
CP-violating 2HDM~\cite{Ginzburg:2002wt} if they have different
unrelated complex phases. In the following, we adopt the conventions
of~\cite{Fontes:2014xva} for the description of the C2HDM. Also, the
phases of the VEVs of the neutral
components of the two Higgs doublets after electroweak symmetry
breaking (EWSB) can in principle be complex in the C2HDM. Without loss of
generality we set them to zero, as they can be removed by a basis
change~\cite{Ginzburg:2002wt}. Expanding the Higgs doublets $\Phi_j$ around
their respective VEVs $v_j$ ($j=1,2$) after EWSB, they can be written as
\beq
\Phi_1 = \begin{pmatrix} \phi_1^+ \\ \frac{v_1+\rho_1 + i \eta_1}{\sqrt{2}} 
\end{pmatrix}  \quad \mbox{and} \quad
\Phi_2 = \begin{pmatrix} \phi_2^+ \\ \frac{v_2+\rho_2 + i \eta_2}{\sqrt{2}} 
\end{pmatrix} \, ,
\eeq
where the $\phi_j^+$ denote the complex charged fields, and $\rho_j$
and $\eta_j$ the neutral CP-even and CP-odd fields, respectively. The
VEVs are related to the SM VEV $v\approx 246$~GeV through
$v_1^2+v_2^2=v^2$, and their ratio is parametrised by the mixing angle
$\beta$,
\beq
\tan\beta \equiv t_\beta = \frac{v_2}{v_1} \;.
\eeq
The mass parameters $m_{11}^2$ and $m_{22}^2$ of the Higgs potential
can be eliminated in favour of $v_1$ and $v_2$ by exploiting the
minimum conditions of the potential that require that its minimum is
given by $\langle \Phi_j \rangle = (0,v_j/\sqrt{2})^T$. They also
relate the imaginary parts of $m_{12}^2$ and $\lambda_5$ and thus fix
one of the ten Higgs potential parameters. Applying the orthogonal
rotation matrix $R$ to the neutral components of the interaction
basis, $\rho_{1,2}$ and $\rho_3 \equiv (1/\sqrt{2}) (-\sin\beta \eta_1
+ \cos\beta \eta_2)$, we obtain the neutral Higgs mass eigenstates $H_i$
($i=1,2,3$), 
\beq
\left( \begin{array}{c} H_1\\ H_2\\ H_3 \end{array} \right) = R
\left( \begin{array}{c} \rho_1 \\ \rho_2 \\ \rho_3 \end{array} \right) \, .
\label{eq:rotc2hdm}
\eeq
Note, that the field $\rho_3$ is equal to the CP-odd component of the
second Higgs doublet in the Higgs
basis~\cite{Lavoura:1994fv,Botella:1994cs}. The matrix $R$
diagonalizes the mass matrix ${\cal M}$ of the neutral states,  
\beq
R\, {\cal M}^2\, R^T = \textrm{diag} \left(m_{H_1}^2, m_{H_2}^2,
  m_{H_3}^2 \right) \;.
\eeq
The $m_{H_i}$ denote the masses of the neutral Higgs bosons and are
ordered by ascending mass, $m_{H_1} \leq m_{H_2} \leq m_{H_3}$. 
Having a mixing of three Higgs states as in the CxSM
  the mixing matrix $R$ can be parametrised in terms of the mixing angles
$\alpha_i$ in the same way as in the CxSM, {\it cf.}~Eq.~(\ref{eq:rotsinglet}).
The Higgs sector of the C2HDM can be described by nine independent
parameters, that we choose to be~\cite{ElKaffas:2007rq}
\begin{equation}
v \;, \quad t_\beta \;, \quad \alpha_{1,2,3}
\;, \quad m_{H_i} \;, \quad m_{H_j} \;, \quad m_{H^\pm} \;, \quad 
\mbox{Re}(m_{12}^2) \;.
\label{eq:2hdminputset}
\end{equation}
Here $m_{H_i}$ and $m_{H_j}$ denote any of the three neutral Higgs
boson masses, while the third mass is not independent and  is calculated
from the other parameters~\cite{ElKaffas:2007rq}. 
The triple Higgs couplings are found from the potential,
\begin{multline}
V={1\over 3!}\sum_{i=1}^3 \lambda_{H_iH_iH_i }H_i^3 
  +{1\over 2} \sum_{i=1}^3\sum_{j=i+1}^3\lambda_{H_i H_i H_j}H_i^2H_j  
\\ 
+ \lambda_{H_1H_2H_3} H_1 H_2 H_3 \, .
\end{multline}
Further details, and
in particular all Higgs couplings of the C2HDM, can be found
in~\cite{Fontes:2017zfn}.\footnote{The tri-linear Higgs couplings, which are quite complicated,
can also be found analytically at \url{http://porthos.tecnico.ulisboa.pt/arXiv/C2HDM}.}

\subsection{The C2HDM Scan}
\label{sec:c2hdmscan}
In this work, for simplicity we only consider the
  C2HDM type 1 (T1) and type 2 (T2), motivated by the fact that they
  cover to a large extent the phenomenological effects to be expected in the C2HDM. 
In order to find valid C2DHM points for our investigations, and
to define benchmark points\footnote{For benchmarks for double Higgs
production in the 2HDM, see {\it{e.g.}}~Refs. \cite{Haber:2015pua,Baglio:2014nea}.} that have not been
excluded yet\footnote{See Ref.~\cite{Basler:2018dac}, for our recently
proposed benchmark points for C2HDM di-Higgs production.}, we
used {\tt ScannerS} to perform a
scan in the C2HDM parameter space. 
As in the CxSM, {\tt ScannerS} checks for the
  theoretical constraints on the C2HDM Higgs potential, and it uses the
  tree-level discriminant of~\cite{Ivanov:2015nea} to enforce 
  the electroweak vacuum to be the global minimum of the tree-level Higgs potential
We require the mass of
the lightest Higgs boson, that is identified with the SM-like one and
denoted by $h$, to be $m_h=125.09$~GeV~\cite{Aad:2015zhl}.
In Tab.~\ref{tab:c2hdmscan} we summarise the ranges of
the other scan parameters. Note that the third
  neutral Higgs boson mass $m_{H_j \ne H_i,h}$ is calculated
  from the other input values and forced to lie in the interval given
  in Tab.~\ref{tab:c2hdmscan}. In order to circumvent degenerate Higgs
  signals, we additionally impose $m_{H_{i,j} \ne h}$ to be 5 GeV away
  from 125~GeV. The SM input parameters are chosen as in the scan for the CxSM. 
\begin{table}
\begin{center}
\begin{tabular}{l|c|c|c|c|c} \toprule
& $t_\beta$ & $\alpha_{1,2,3}$ & $\mbox{Re}(m_{12}^2)$ [TeV$^2$] & $m_{H^\pm}$ [TeV] & $m_{H_{i,j}\ne h}$ [TeV] \\ \hline
min & 0.8 & $-\frac{\pi}{2}$ & 0 & 0.15/0.59 & {{0.125}} \\
max & 20 & $\frac{\pi}{2}$ & 0.5 & 1.5 & 1.5 \\ \bottomrule
\end{tabular}
\caption{C2HDM scan: All parameters are varied
  independently between the given minimum and maximum values. The two
  minimum values of the charged Higgs mass range refer to the scan in
  the C2HDM T1 and T2, respectively. For more details, see text. 
\label{tab:c2hdmscan}}
\end{center}
\end{table}
In  our scan we neglect parameter points with $\mbox{Re}
(m_{12}^2) <0$, as they are extremely rare. We check all parameter
points at the 2$\sigma$ exclusion level in the $m_{H^\pm}-\tan\beta$
plane for  compatibility with the flavour constraints on
$R_b$~\cite{Haber:1999zh,Deschamps:2009rh} and $B \to X_s
\gamma$~\cite{Deschamps:2009rh,Mahmoudi:2009zx,Hermann:2012fc,Misiak:2015xwa,Misiak:2017bgg}
Applying the results of~\cite{Misiak:2017bgg} we require $m_{H^\pm}$
to be above 590~GeV in the C2HDM T2. In the C2HDM T1, on the other
hand, the bound is much weaker and depends more strongly on
$\tan\beta$. Our retained parameter points are put in agreement
with the electroweak precision data by demanding $2\sigma$
compatibility with the SM fit~\cite{Baak:2014ora} of the oblique
parameters $S$, $T$ and $U$, including the full
correlation among the three parameters. The necessary 2HDM formulae are given
in~\cite{Branco:2011iw,Dawson:2017jja}. For the check of
the compatibility with the Higgs data we proceeded as in the CxSM,
with the difference that we obtained the here necessary branching
ratios from the C2HDM implementation {\tt 
  C2HDM\_HDECAY}~\cite{Fontes:2017zfn} in {\tt
  HDECAY}~\cite{Djouadi:1997yw,Djouadi:2018xqq}.
Further details, can be found in~\cite{Fontes:2017zfn,Muhlleitner:2017dkd}. 

Since we work in the C2HDM, we also have to check for agreement with
the measurements of the electric dipole moment (EDM), with the strongest constraint
originating from the electron EDM~\cite{Inoue:2014nva}. We take the
experimental limit given by the ACME
collaboration~\cite{Andreev:2018ayy}.
As shown explicitly in Figs. 5 and 6 of Ref. \cite{Chen:2015gaa}, the limits from
the neutron EDM are weaker than those from the electron EDM.
 Like for the
  CxSM we also checked if the final scenarios induce a strong first order
  phase transition \cite{Basler:2018cwe,Basler:2017uxn}. Also here we
  found that for none of them this is the case. 

\section{Interference effects: top vs. di-Higgs final states}
\label{sec:res}
\subsection{Setup}

Based on the scan detailed in Sec.~\ref{sec:models},  we implement the $pp \to H_i \to t \bar t$ and $pp\to H_i \to h h$ resonant
amplitudes into
{\sc{Vbfnlo}}~\cite{Arnold:2008rz,Arnold:2011wj,Arnold:2012xn,Baglio:2014uba}, 
where
$H_i$ denotes any of the non-SM-like heavy Higgs bosons of the CxSM or
C2HDM, respectively. For the parameter regions investigated here the
main production channel is given by gluon fusion. The one-loop
(leading order) computation uses
{\sc{FormCalc/LoopTools}}~\cite{Hahn:1999mt,Hahn:2001rv}. 
Various cross checks
against {\sc{MadGraph}}~\cite{Alwall:2014hca} and other
results~\cite{Spira:1995mt,Spira:1996if,Anastasiou:2009kn,Dittmaier:2011ti}
have been carried out.
We do not include $b$ quark loops throughout as they are
  negligible for the parameter regions studied in this work.\footnote{Specifically, for the T2 scenario
  we always observe $\tan\beta \sim 1$ while for T1 most points show $\tan\beta \simeq 7$.}
  
\begin{figure}[!b]
\includegraphics[width=0.24\textwidth]{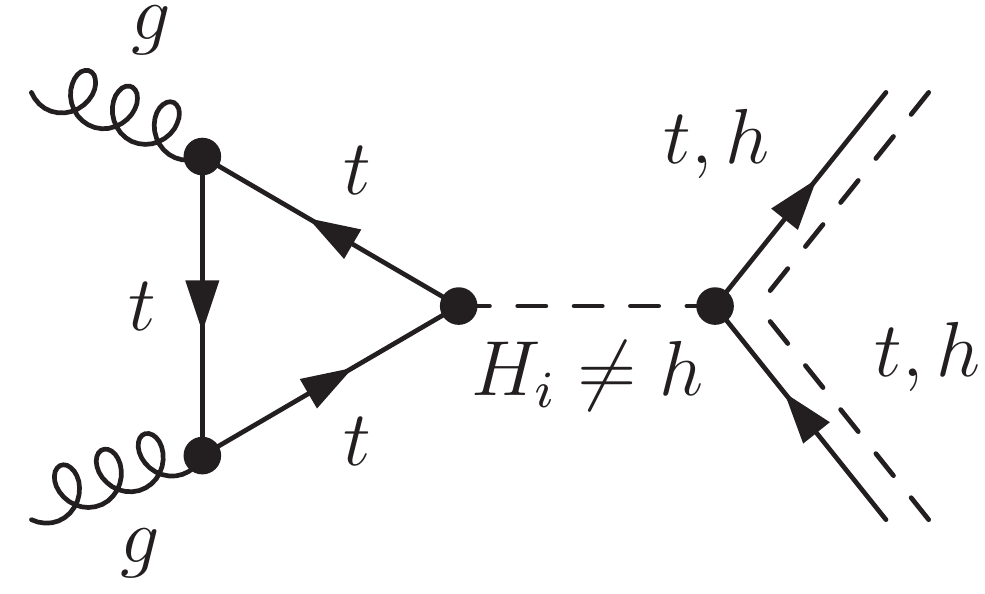}
\caption{ \label{fig:sigdiag} Representative signal diagram
  contributing to $t\bar t $ and $hh$ resonance searches. $h$ denotes the light SM-like state with $m_h\simeq 125~\text{GeV}$,  while $H_i$ denotes the remaining heavy Higgs bosons that arise in the C2HDM and CxSM.}
\end{figure}

\begin{figure*}[!t]
\includegraphics[width=0.69\textwidth]{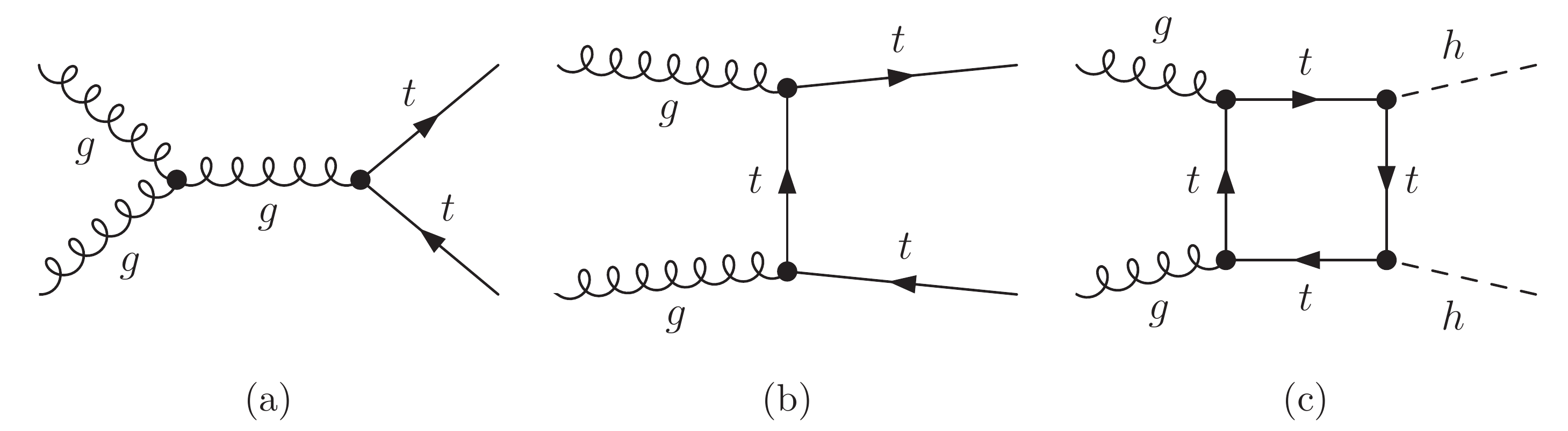}
\caption{ \label{fig:bkgdiag} Representative non-resonant
  ``background'' diagrams contributing to $pp\to t\bar t $ (a,b) and
  $pp \to hh$ (c) searches (different fermion flows are understood
  implicitly). The off-shell $h$-induced
    background contribution derives 
    from graphs shown in Fig.~\ref{fig:sigdiag} with an off-shell $h$
    running in the $s$-channel.}
\end{figure*}
\begin{figure*}[!t]
\subfigure[\label{fig:2hdmt1}]{\includegraphics[width=0.4\textwidth]{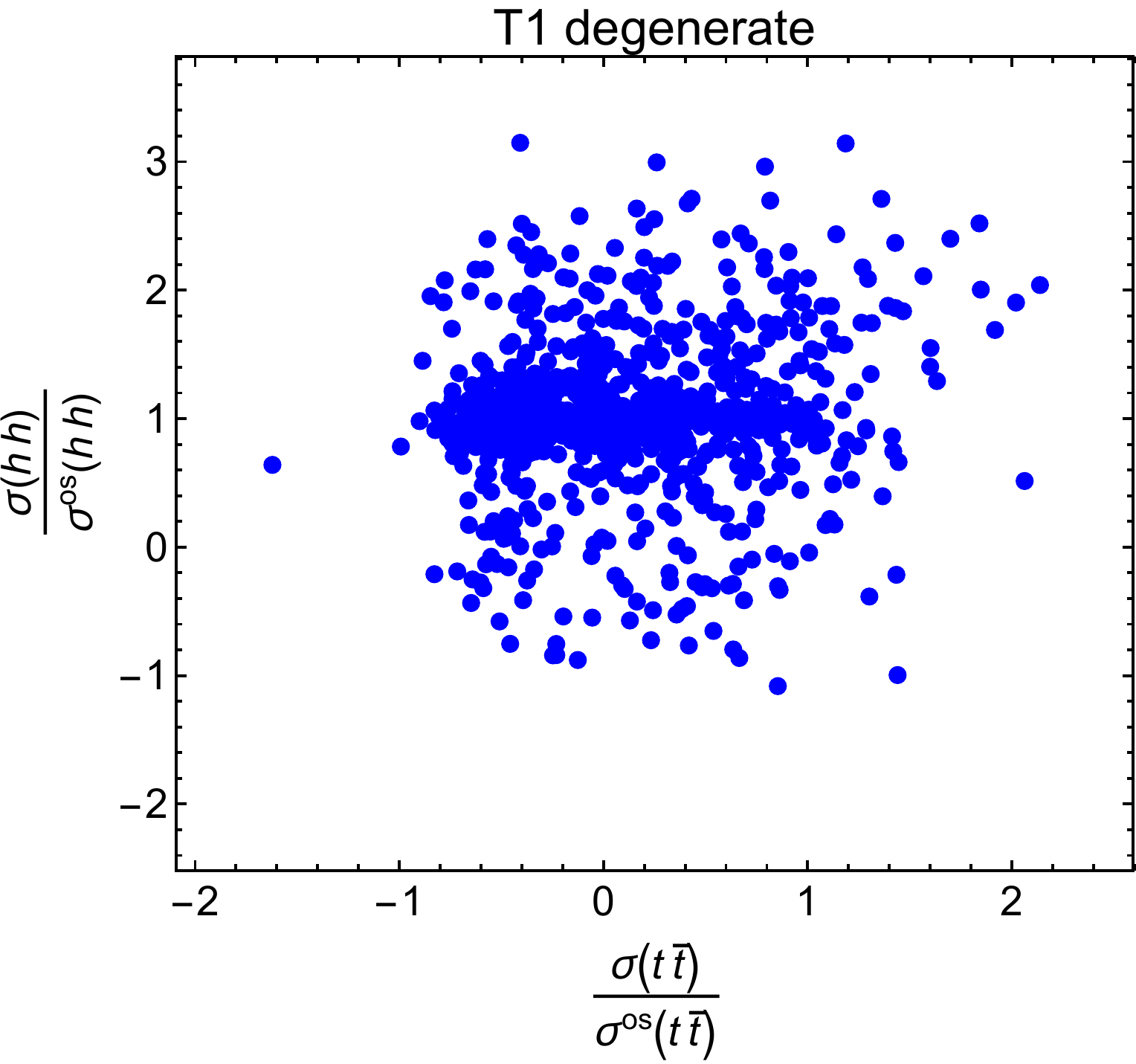}}
\qquad
\subfigure[\label{fig:2hdmt2}]{\includegraphics[width=0.4\textwidth]{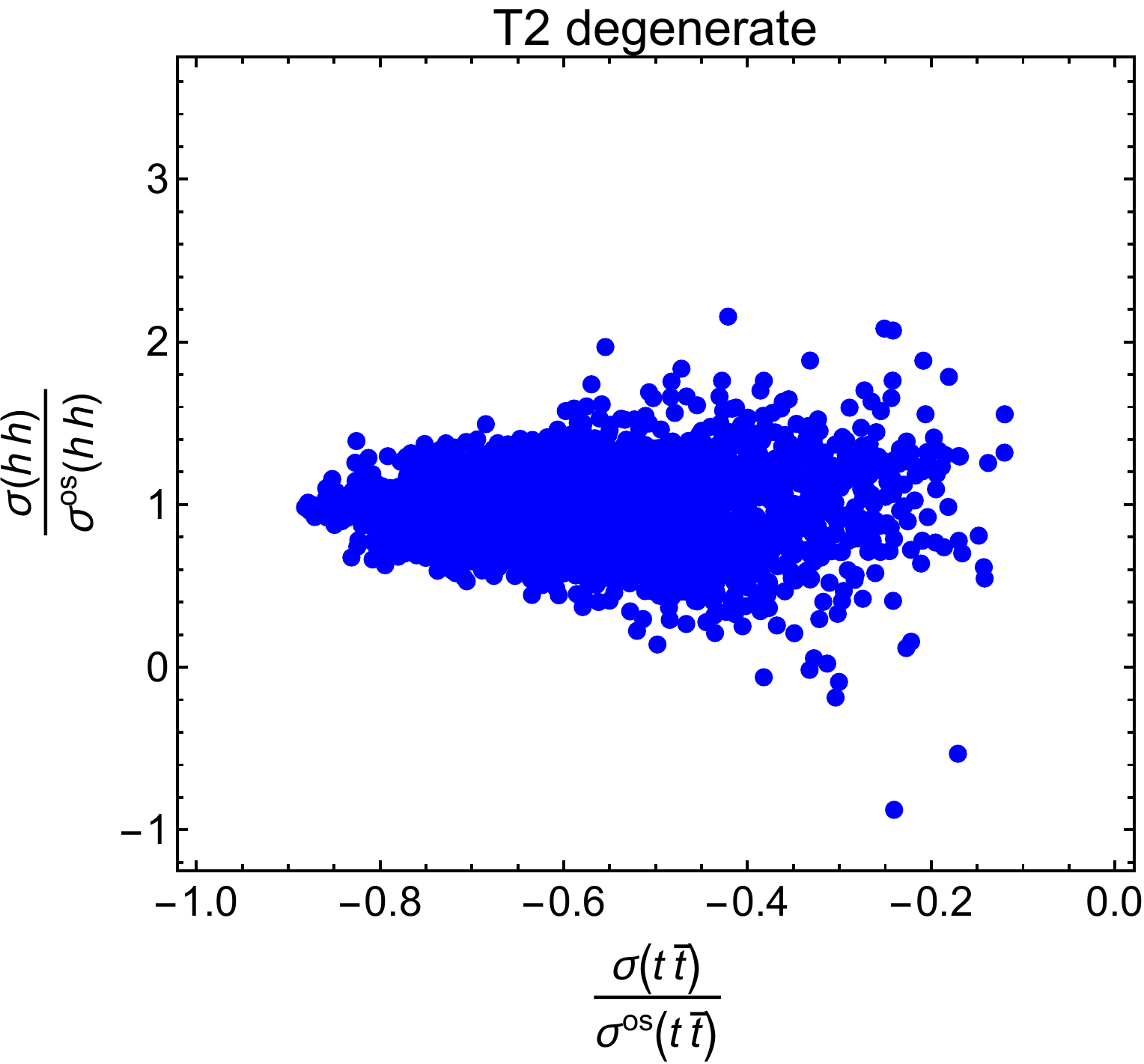}}
\vspace*{-0.1cm}
\caption{Ratio of signal+interference cross section $\sigma$ and OS
  cross-section $\sigma^{\text{os}}$ (for definition, see text) in 
  $pp\to hh$ and $pp\to t\bar{t}$ for degenerate
    non-SM-like Higgs states. Points are pre-selected to have
  resonance cross sections of at least 170 fb at LO in the $t\bar t$
  and 8 fb in the $hh$ channels. Left: 2HDM type 1,
    right: 2HDM, type 2. \label{fig:correlation}} 
\end{figure*}
\begin{figure*}[!b]
\subfigure[~]{\includegraphics[width=0.48\textwidth]{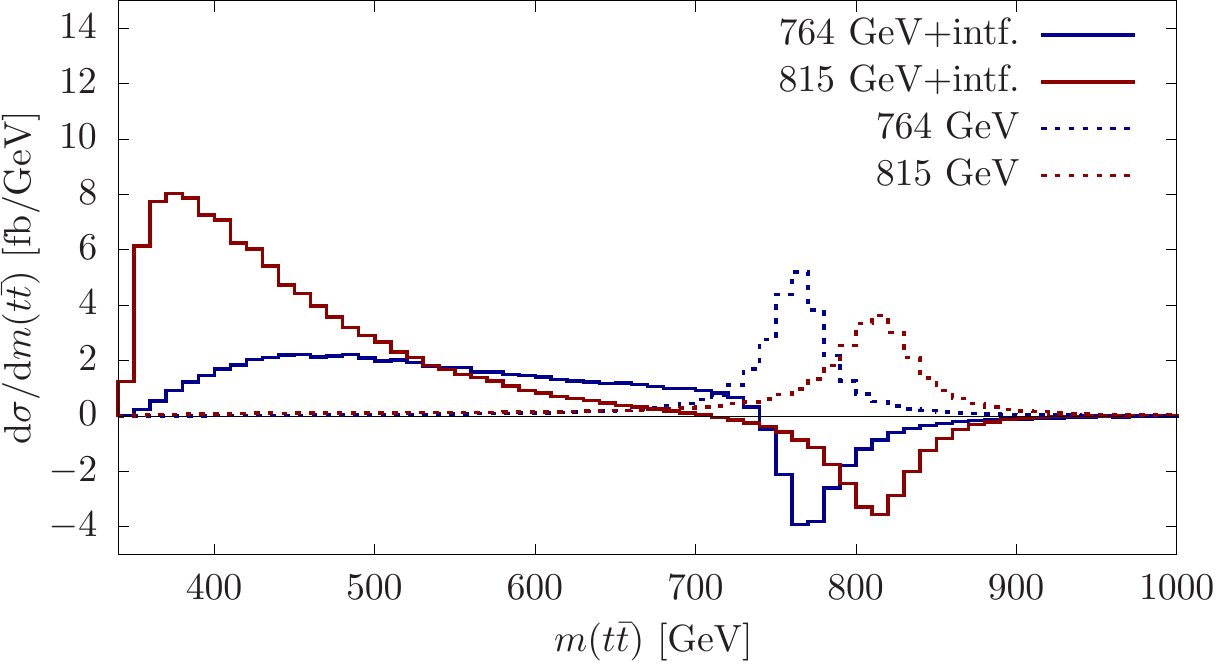}} \hfill
\subfigure[~]{\includegraphics[width=0.48\textwidth]{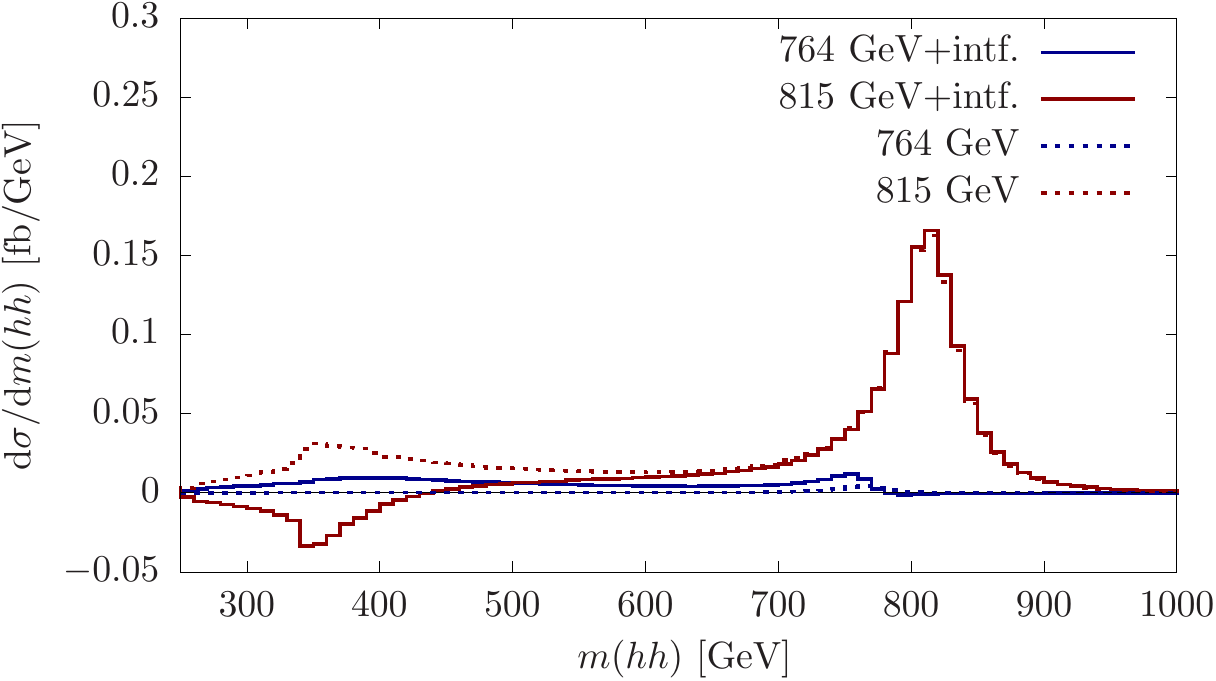}}
\caption{Comparison of the invariant mass distribution
    for $pp \to t\bar t$ (left) and $pp\to h h$ (right) at 13 TeV at
  LO for the different states $H_{i}\neq h$ (blue: $H_i=H_2$, red: $H_i=H_3$). We show the signal $gg\to t\bar t $ 
and $gg\to h h$ production following Eq.~\eqref{eq:sigdef} as dashed lines. The interference-corrected cross sections, Eq.~\eqref{eq:totdef}, are depicted as solid lines. The spectra arise from the parameter point {\texttt{BP1}}, see Tab.~\ref{tab:bench}. \label{fig:bp1164}  }
\end{figure*}

We select one state $H_i$, defined as
the signal, and compute the squared amplitude for the $gg \to H_i \to t\bar t / hh$ process:
\begin{equation}
\label{eq:sigdef}
\hbox{d}\sigma^\text{os}_i \sim |{\cal{M}}_{\text{sig}} (gg\to H_i \to X\bar{X})|^2 \,,\quad X=t,h\,,
\end{equation}
where ${\cal{M}}$ is the signal amplitude given by the $s$-channel one-loop diagrams shown in Fig.~\ref{fig:sigdiag}. This cross section can be understood as the
on-shell cross section that one would obtain from
$\sigma$-times-branching ratio estimates. To obtain these cross
sections and put them  in relation to interference effects,  we
integrate the cross sections within 
\begin{equation}
\label{eq:intrange}
|m(t\bar t/hh)-m_{H_i}| < 2 \,\Gamma_{H_i}\,.
\end{equation}
We keep track of the interference effects
with the SM ``background'' and BSM signal. The former is given by
continuum $gg\to t\bar t$ production, Fig.~\ref{fig:bkgdiag}, for the
$t\bar{t}$ final state, and by box, Fig.~\ref{fig:bkgdiag}, and off-shell
$h$-induced $gg \to hh$ contributions for the $hh$ final
state. The latter derives from the competing $gg\to H_{j\neq i}\to hh$
diagrams, Fig.~\ref{fig:sigdiag}.
This gives rise to an estimate of the observed cross section in the
presence of interference effects: 
\begin{multline}
\label{eq:totdef}
\hbox{d}\sigma_i \sim|{\cal{M}}_{\text{sig}} (gg\to H_i \to
X\bar{X})|^2 \\+ 2\,\hbox{Re} \left\{ {\cal{M}}_{\text{sig}}
  {\cal{M}}^\ast_\text{bkg}(H_{j\neq i},\text{cont.}) \right\}\,,
\end{multline}
where ``cont.'' stands for the continuum $t\bar t $ or $hh$ ``background'' and (off-shell) $H_{j\neq i}$ contributions as mentioned above, including the SM-like $h$.

The scans described in the previous section show that there are viable parameter choices with the tendency to produce quasi-degenerate mass spectra in the C2HDM when both $t\bar t$ and $hh$ decay channels
are open. 
We define the two non-SM states as ``degenerate'' when their mass splitting is less than 10\% of the heavy scalar's mass. This accounts for
most of the parameter points that are described in Sec.~\ref{sec:models}. 

For parameter points that have very small cross sections
in either of the two channels, interference effects when considered in relation to 
the on-shell signal definition can be very large, however in this case they have little phenomenological importance. We therefore filter our results with some minimum
cross section requirements for both $pp\to t  \bar t$ and $pp \to
hh$. For $pp\to t\bar t$ we require at least 170 fb before the
inclusion of $K$ factors, for $pp\to hh$ we demand at least
$8$~fb. This amounts to about
${\cal{O}}(0.5~\text{pb})$~\cite{Anastasiou:2014lda,Anastasiou:2014vaa}
when higher-order corrections are included for $t \bar t$ final states
and $\simeq 16~\text{fb}$ for
$hh$ production~\cite{Borowka:2016ehy,Borowka:2016ypz,Heinrich:2017kxx,Baglio:2018lrj,Davies:2018qvx,Bonciani:2018omm,Chen:2014ask,Grober:2017gut}.

\begin{figure*}[!t]
\subfigure[~]{\includegraphics[width=0.48\textwidth]{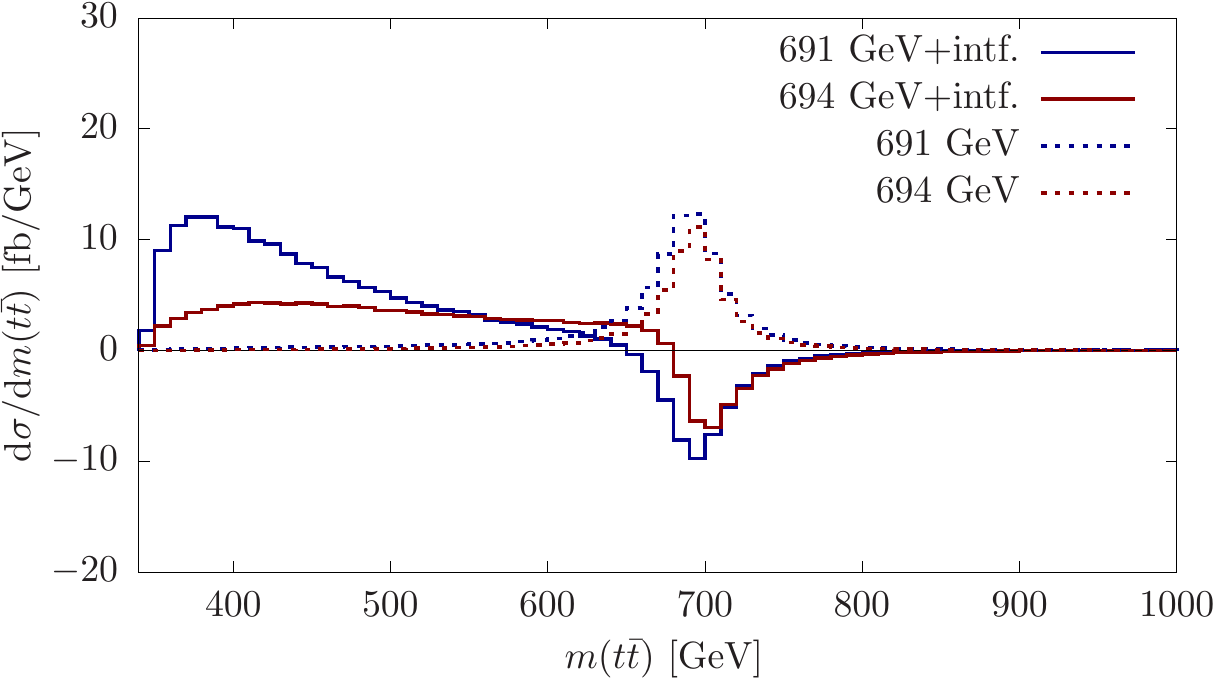}  }\hfill
\subfigure[~]{\includegraphics[width=0.48\textwidth]{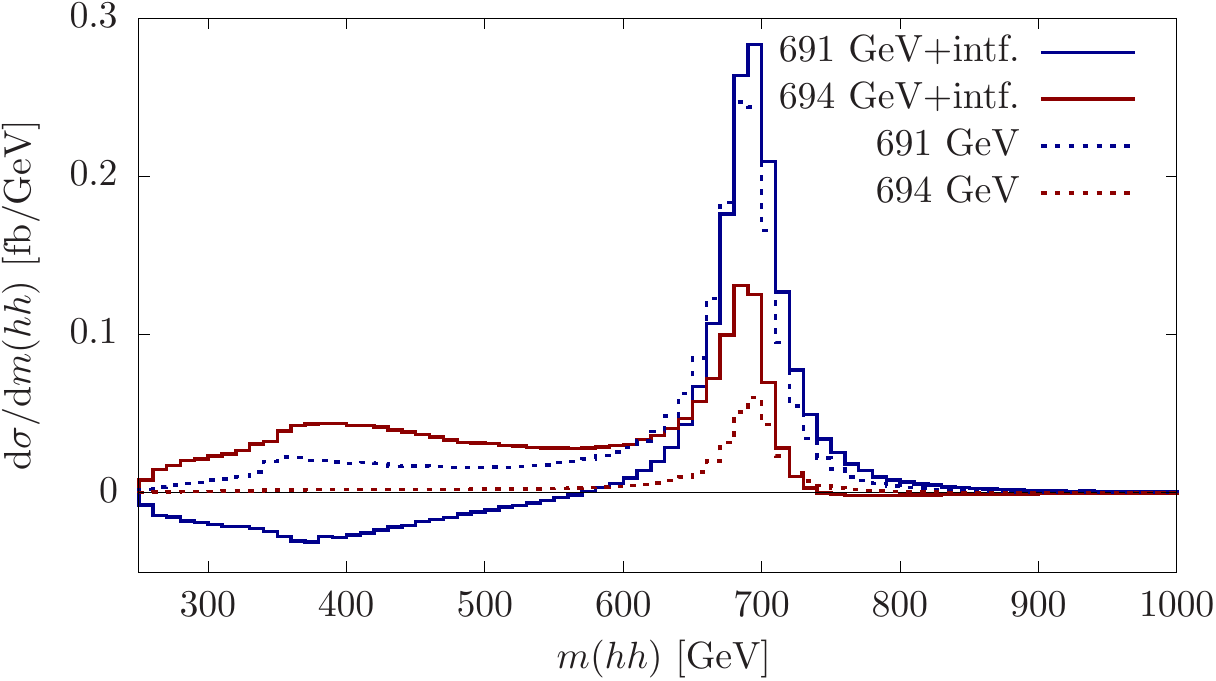}}
\caption{Comparison of the invariant mass distribution
    for $pp \to t\bar t$ (left) and $pp\to h h$ (right) at 13 TeV at
  LO for the different states $H_{i}\neq h$ (blue: $H_i=H_2$, red: $H_i=H_3$). We show the signal $gg\to t\bar t $ 
and $gg\to h h$ production following Eq.~\eqref{eq:sigdef} as dashed lines. The interference-corrected cross sections, Eq.~\eqref{eq:totdef}, are depicted as solid lines. The spectra arise from the parameter point {\texttt{BP2}}, see Tab.~\ref{tab:bench}.\label{fig:bp17783}}
\end{figure*}
\begin{figure*}[!t]
\subfigure[~]{\includegraphics[width=0.48\textwidth]{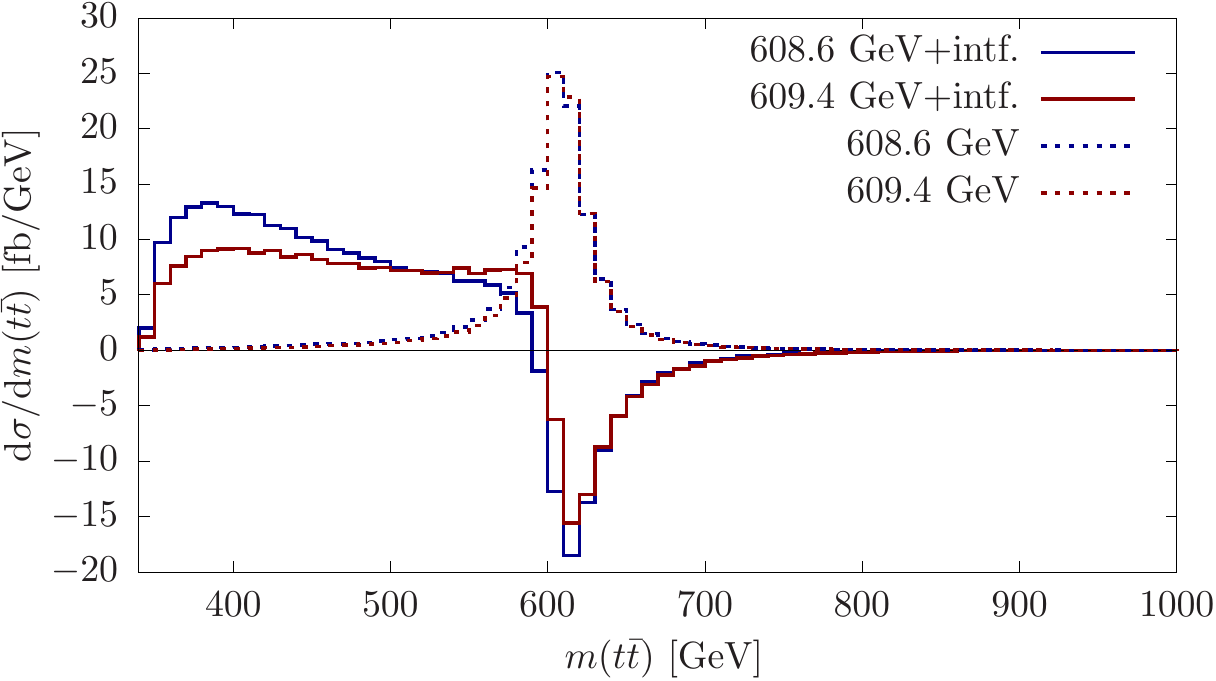} } \hfill
\subfigure[~]{\includegraphics[width=0.48\textwidth]{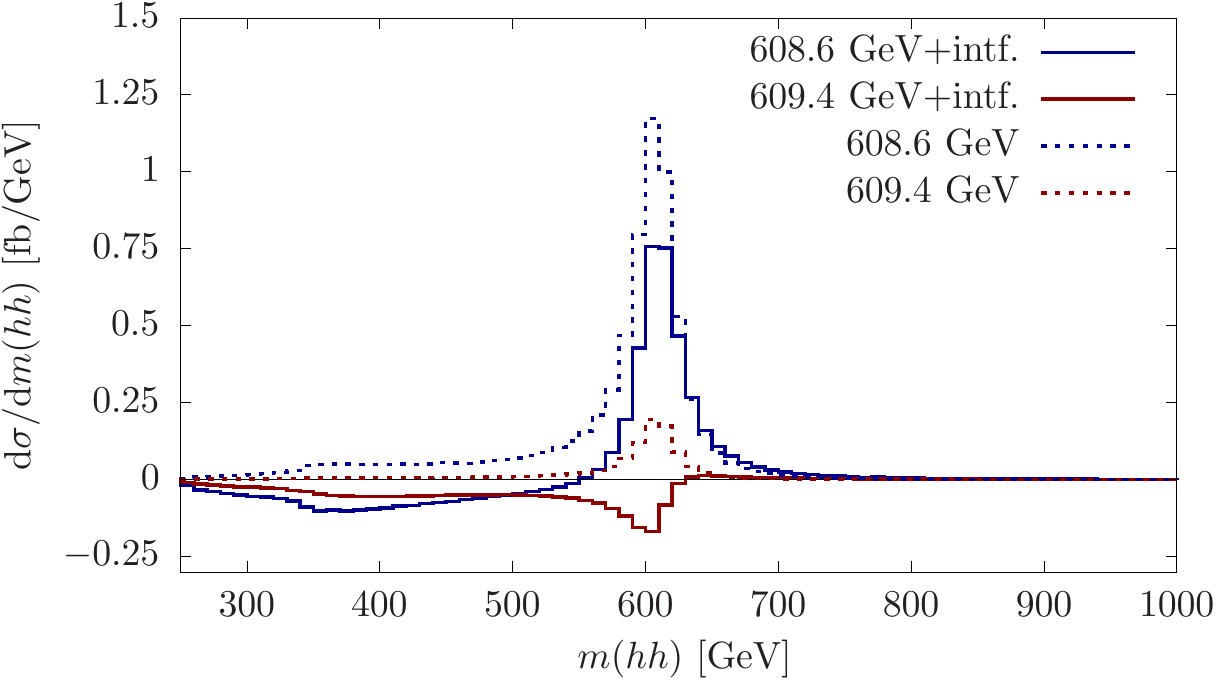}}
\caption{Comparison of the invariant mass distribution
    for $pp \to t\bar t$ (left) and $pp\to h h$ (right) at 13 TeV at
  LO for the different states $H_{i}\neq h$ (blue: $H_i=H_2$, red: $H_i=H_3$). We show the signal $gg\to t\bar t $ 
and $gg\to h h$ production following Eq.~\eqref{eq:sigdef} as dashed lines. The interference-corrected cross sections, Eq.~\eqref{eq:totdef}, are depicted as solid lines. The spectra arise from the parameter point {\texttt{BP3}}, see Tab.~\ref{tab:bench}.\label{fig:bp10544}  }
\end{figure*}
\begin{figure*}[!t]
\subfigure[~]{\includegraphics[width=0.48\textwidth]{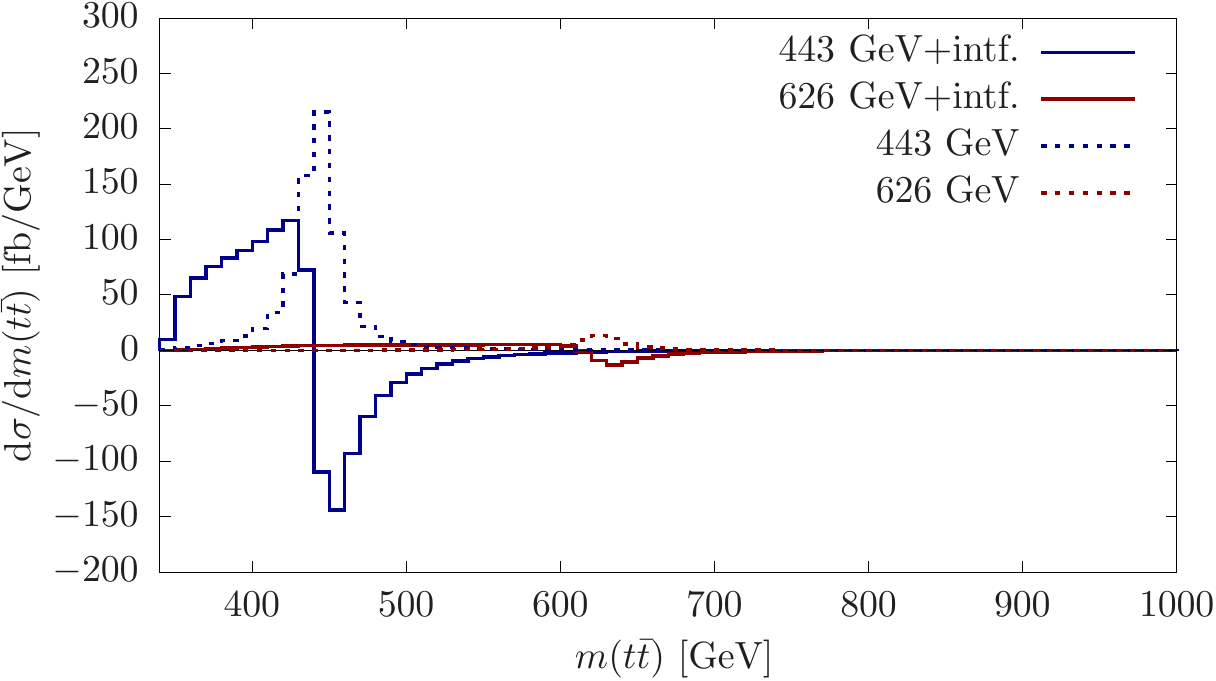} }\hfill
\subfigure[~]{\includegraphics[width=0.48\textwidth]{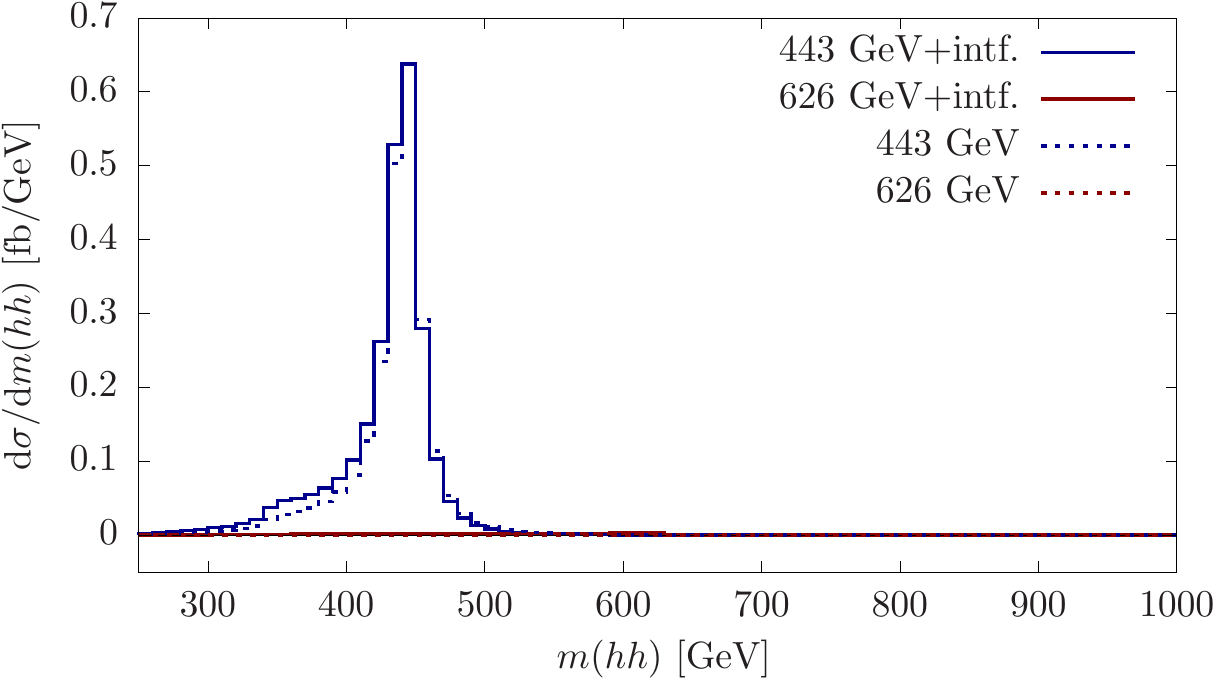}}
\caption{Comparison of the invariant mass distribution
    for $pp \to t\bar t$ (left) and $pp\to h h$ (right) at 13 TeV at
  LO for the different states $H_{i}\neq h$ (blue: $H_i=H_2$, red: $H_i=H_3$). We show the signal $gg\to t\bar t $ 
and $gg\to h h$ production following Eq.~\eqref{eq:sigdef} as dashed lines. The interference-corrected cross sections, Eq.~\eqref{eq:totdef}, are depicted as solid lines. The spectra arise from the parameter point {\texttt{BP4}}, see Tab.~\ref{tab:bench}.\label{fig:bp1019}  }
\end{figure*}

\subsection{Results and Discussion}
\label{sec:res}
\subsubsection{The C2HDM}
In order to investigate the effects from interferences for the $hh$
and $t\bar{t}$ final states, we introduce the
ratio of the signal plus interference cross section $\sigma$ (defined in
Eq.~(\ref{eq:totdef})) and the signal cross section
$\sigma^{\text{os}}$ (defined in Eq.~\eqref{eq:sigdef} for the
requirement Eq.~\eqref{eq:intrange}), {\it{i.e.}}
\beq
R(xx) = \frac{\sigma (xx)}{\sigma^{\text{os}} (xx)} \;, \quad xx=hh, t\bar{t}\;.
\eeq
In Fig.~\ref{fig:2hdmt1} we show $R(hh)$ versus $R(t\bar{t})$ for
the C2HDM type 1 for degenerate non-SM-like Higgs
  states, {\it i.e.}~states whose masses differ by less than 10\%. As can  
be inferred from the figure, there is a broad range of possible
phenomenological outcomes. We can have a large enhancement or
suppression of the $H_i\to t\bar t$ signal while the $hh$ rate can be
either enhanced or reduced. Points with large constructive
interference effects in the $t\bar{t}$ final state are likely to be
constrained through $pp \to t\bar t $ measurements. We also obtain
parameter points for which interference effects decrease the search
potential in both the $t\bar t $ and $hh$ channels. Having
simultaneous contributions from signal-signal ({\it{i.e.}} interference
  between the two $s$-channel $H_{i}\neq h$ contributions) and
signal-background interference for the resonance masses not too far
away from each other, both effects contribute when we obtain a
simultaneous enhancement in the $t\bar t$ and $hh$ rates. Points
that show this correlation are clustered around the $t\bar t$
threshold,  $m_{H_i}\simeq 350~\text{GeV}$,  with small widths
$\Gamma_{H_i}/m_{H_i} \simeq 10^{-3}$: In this region
interference effects with background contributions are particularly
large. As this effect tends to be destructive above the resonance
mass (see also \cite{Aaboud:2017hnm} or below), widening the
mass range that we use to define the signal Eq.~\eqref{eq:intrange}
can decrease the $\sigma(t\bar t)/\sigma^{\text{os}}(t\bar t)$
value. This can be relevant when the mass resolution is imperfect in the
analyses and should be kept track  of in the experimental
investigation. Above the threshold, when $H_i\to
t\bar t$ is kinematically accessible, so is $H_i\to hh$. The
enhancement in $hh$ is then a combination of signal-background (see
also \cite{Dawson:2015haa}) and signal-signal interference, with the
latter playing the dominant role for the parameter choices studied
here. 

Particularly interesting from a double Higgs discovery perspective,
however,  is the tendency of the $H_i\to t\bar t$ signals to be
reduced, with large constructive interference effects present in
$H_i\to hh$. As the box-graph ``backgrounds'' in the $pp\to hh$ case
decrease with the centre-of-mass energy, these effects are
predominantly due to signal-signal interference for overlapping
Breit-Wigner distributions when the mass spectra are
quasi-degenerate. 
This is a key implication of the scan as detailed in
Sec.~\ref{sec:models}: While signal-background interference decreases
the sensitivity in the $t\bar t$ channels, which are known to be the
most constraining channels in C2HDM sensitivity extrapolations (see
{\it{e.g.}}~\cite{Basler:2018dac}), these effects can be correlated
with large signal-signal interference effects in the di-Higgs
modes. Not only is the $\sigma$-times-branching ratio estimate typically
inadequate in both decay modes, but di-Higgs final states can become
dominant search modes for new physics in the context of the 2HDMs as
discussed above.

\begin{table}[!b]
\begin{tabular}{|| lcccc ||}
\hline
\toprule
{} &      {\texttt{BP1}} &    {\texttt{BP2}} &   {\texttt{BP3}} &  {\texttt{BP4}} \\
\midrule
\hline
$m_{H_1}\,[\mathrm{GeV}]$           &  125.090 &  125.090 &  125.090 &  125.090 \\
$m_{H_2}\,[\mathrm{GeV}]$           &  764.044 &  691.319 &  608.588 &  442.903 \\
$m_{H_3}\,[\mathrm{GeV}]$           &  814.578 &  694.637 &  609.393 &  626.371 \\
$m_{H^\pm}\,[\mathrm{GeV}]$         &  853.064 &  654.204 &  679.601 &  651.550 \\
\hline
$\alpha_1$                          &    0.746 &    0.766 &    0.818 &    0.736 \\
$\alpha_2$                          &   -0.132 &    0.042 &    0.053 &    0.045 \\
$\alpha_3$                          &   -0.086 &    1.144 &    0.913 &    1.567 \\
$\tan(\beta)$                       &    0.921 &    0.870 &    0.892 &    0.928 \\
$R_{13}^2$                          &    0.017 &    0.002 &    0.003 &    0.002 \\
$R_{23}^2$                          &    0.007 &    0.827 &    0.624 &    0.998 \\
$R_{33}^2$                          &    0.975 &    0.171 &    0.373 &    0.000 \\
\hline
\hline
$\sigma(gg \to H_1)\,[\mathrm{pb}]$ &   45.908 &   49.699 &   53.640 &   43.233 \\
$\sigma(gg \to H_2)\,[\mathrm{pb}]$ &    0.651 &    1.700 &    2.903 &   19.042 \\
$\sigma(gg \to H_3)\,[\mathrm{pb}]$ &    0.637 &    1.284 &    2.670 &    1.899 \\
\hline
\hline
$\lambda_{H_1 H_1 H_1} \,{\mathrm{GeV}]}$             &  -30.633 &  150.815 &  115.626 & -184.173 \\
$\lambda_{H_1 H_1 H_2} \,{[\mathrm{GeV}]} $            &  -49.478 &  253.524 &  305.386 &  -55.652 \\
$\lambda_{H_1 H_1 H_3}\, {[\mathrm{GeV}]}$             & {-448.381}&  {120.882} &
 { -121.714} & { 6.123} \\
\hline
\hline
$\Gamma(H_1) \,[\mathrm{GeV}]$      &    0.004 &    0.004 &    0.004 &    0.004 \\
$\Gamma(H_2) \,[\mathrm{GeV}]$      &   36.623 &   41.150 &   31.551 &   21.580 \\
$\Gamma(H_3) \,[\mathrm{GeV}]$      &   51.865 &   34.787 &   29.057 &   32.449 \\
\hline
\hline
$BR(H_2\to H_1 H_1)$                &    0.001 &    0.021 &    0.044 &    0.003 \\
$BR(H_2\to t\overline{t})$          &    0.936 &    0.962 &    0.922 &    0.990 \\
$BR(H_3\to H_1 H_1)$                &    0.045 &    0.006 &    0.008 &    0.000 \\
$BR(H_3\to t\overline{t})$          &    0.871 &    0.979 &    0.965 &    0.793 \\
\bottomrule
\hline
\end{tabular}
\caption{Parameter points as shown in
  Figs.~\ref{fig:bp1164}--\ref{fig:bp1019}. For details of the
  calculation see Sec.~\ref{sec:c2hdmscan}. All 
points are not constrained by the most recent CMS analysis of
Ref.~\cite{Sirunyan:2019wph}. Note that for the plots we used LO
rates. The benchmarks reflect the experimental status as summarised
in the tool chain detailed in Sec.~\ref{sec:models}.
 \label{tab:bench} }
\end{table}

This point is further highlighted when we consider the C2HDM of type 2
in Fig.~\ref{fig:2hdmt2}.\footnote{As detailed in
  Ref.~\cite{Basler:2017nzu}, an additional requirement on the charged
  Higgs mass $m_{H^\pm}>590$~GeV for type 1 (for type 2
  $m_{H^\pm}>590$~GeV is enforced by the constraint from the weak
  radiative $B$ meson decays \cite{Misiak:2017bgg}) leads to
  qualitative agreement between the type 1 and type 2 models.} Here we
always find a decrease of the expected leading-order rate for the
$t\bar t $ spectrum. While this can be partly compensated for via
large QCD corrections for the signal component\footnote{Note that as
  we typically deal with small width values, issues that relate to the
  precise definition of signal and
  background~\cite{Passarino:2010qk,Goria:2011wa} are numerically
  suppressed and $\sigma$-times-branching ratio extrapolations would be
  justified if there was no interference.}, it is clear that the
straightforward $\sigma$-times-branching ratio approximation will
overestimate the sensitivity dramatically. Again different outcomes
in the correlation between $hh$ and $t\bar{t}$ final states
are possible: signal-signal interference in $H_i\to hh$ can enhance or
decrease the cross section expectation. Note, however, that in our scan
the points with $\sigma(t\bar t)/\sigma^{\text{os}}(t\bar t)\to -1$
merge towards $\sigma(hh)/\sigma^{\text{os}}(hh)\to 1$.  

The reason for this behaviour becomes apparent from
Fig.~\ref{fig:bp1164}: On the one hand, signal-background interference
in the $t\bar t$ mass spectrum remains large due to the large
continuum background contribution even for heavy resonances of around
$800~\text{GeV}$. On the other hand, hierarchies in the trilinear
Higgs self-couplings are directly reflected in different resonance
cross sections, which implies that $pp\to H_2\to hh$ is much more
suppressed than $pp\to H_3\to hh$, which also implies that
signal-signal interference is not relevant for the $H_3\to hh$
  decay with phenomenologically significant cross section. 
For the particular benchmark point {\texttt{BP1}} (see
Tab.~\ref{tab:bench}), that we choose to illustrate this effect in
Fig.~\ref{fig:bp1164}, we find a large signal-background interference
such that we end up with a dip structure in $t\bar t$ with about
$\simeq -200$ fb when integrating Eq.~\eqref{eq:totdef} over the range
of Eq.~\eqref{eq:intrange}.

\begin{figure*}[!t]
\subfigure[\label{fig:sdeg}]{\includegraphics[width=0.4\textwidth]{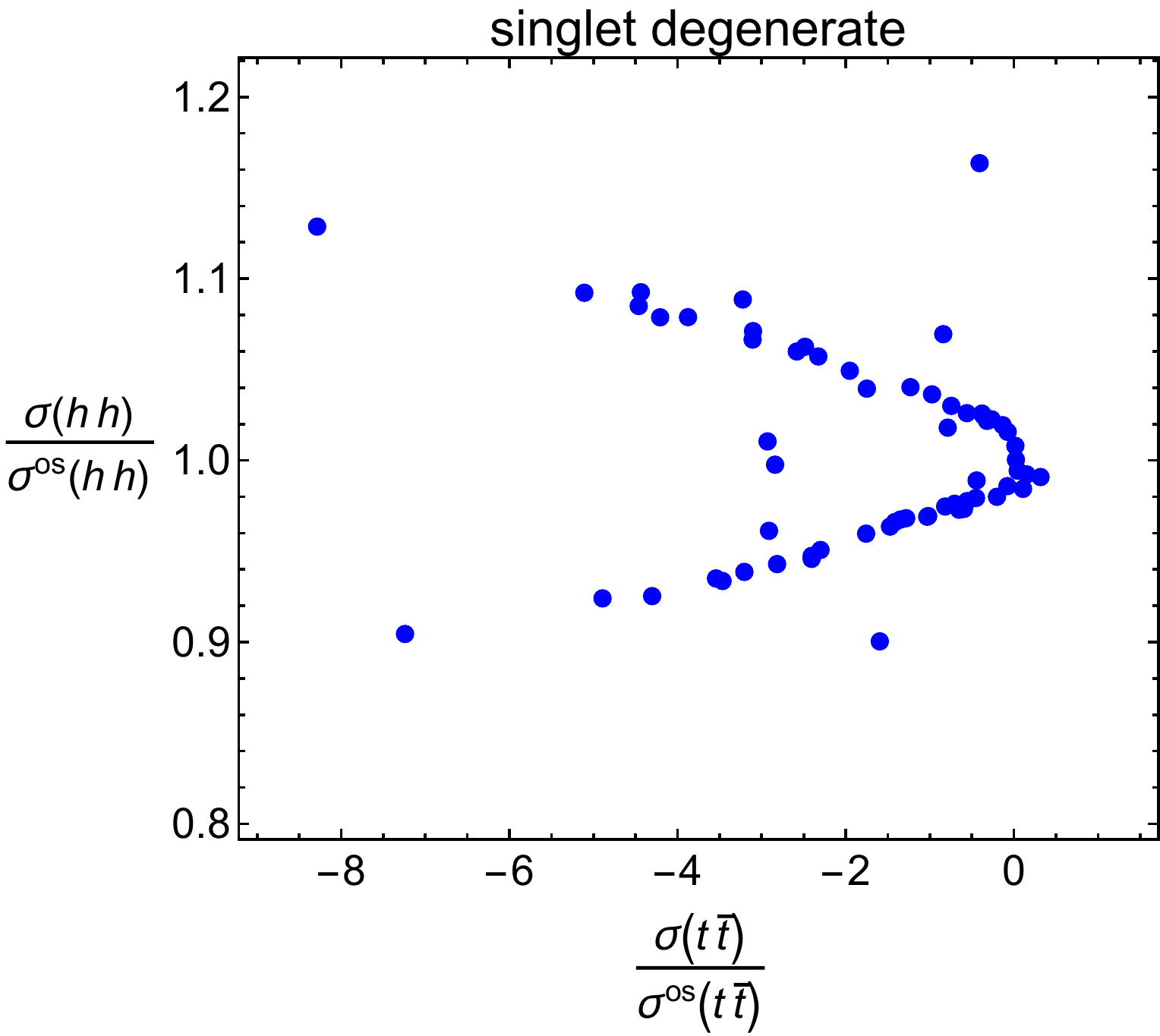}} \qquad
\subfigure[\label{fig:sndeg}]{\includegraphics[width=0.4\textwidth]{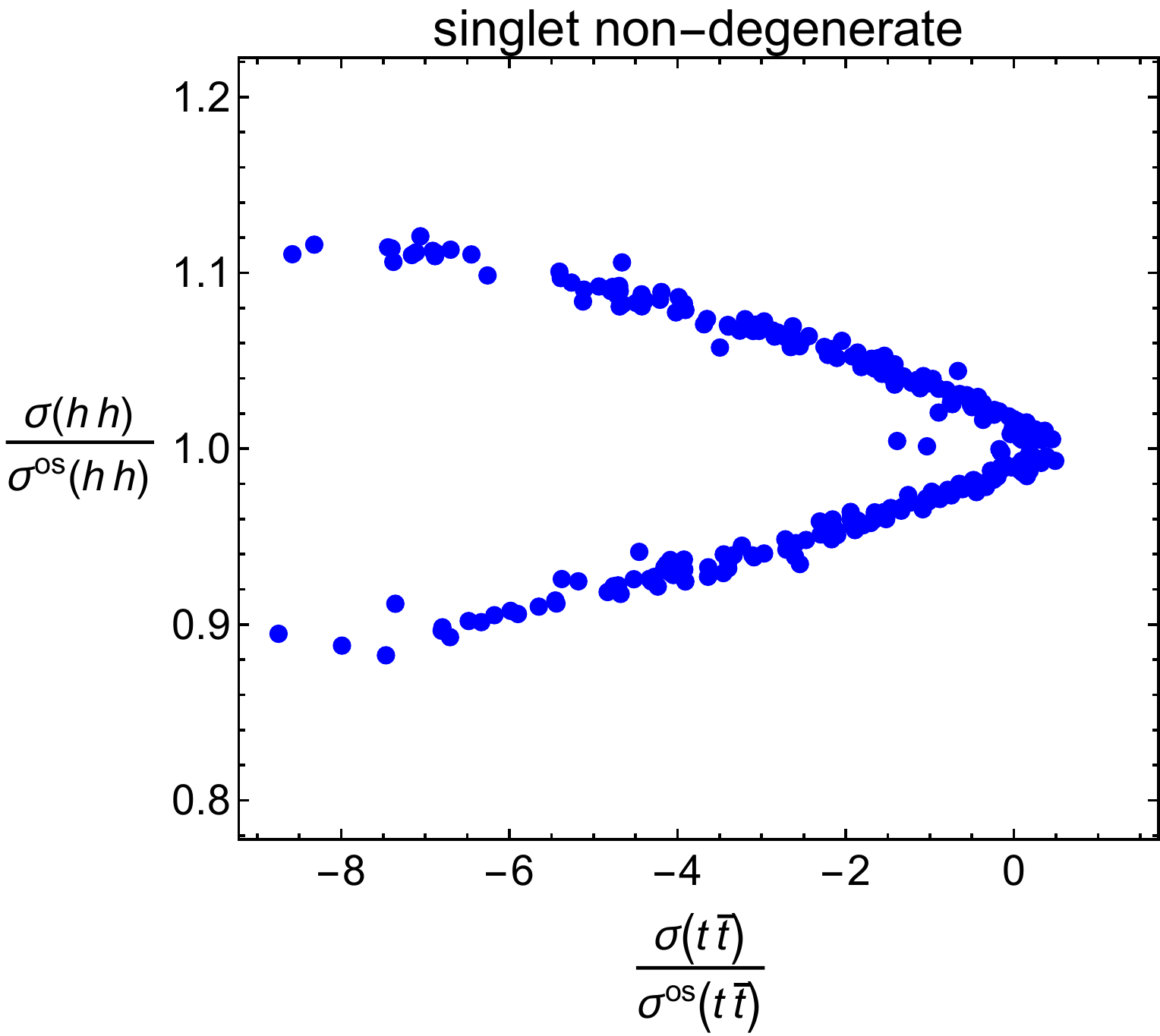}}
\caption{Ratio of signal+interference cross section $\sigma$ and OS
  cross-section $\sigma^{\text{os}}$ in 
  $pp\to hh$ and $pp\to t\bar{t}$ in the CxSM scenario. Points are
pre-selected to have at least 8 fb in the $hh$ channels at leading
order. Due to the nature of this model the absolute values of the
$t\bar t$ final state rates are
phenomenologically irrelevant as they are 
${\cal{O}}({\text{fb}})$. For details see text. Left: degenerate,
right: non-degenerate non-SM-like Higgs bosons. \label{fig:singlet}} 
\end{figure*}

Particularly interesting are again parameter points for which $t\bar
t$ is decreased relative to the on-shell expectation while $hh$ is
enhanced. 
An example of such an outcome is {\texttt{BP2}} (see 
Tab.~\ref{tab:bench}). As can be inferred from
  Fig.~\ref{fig:bp17783}, we find a large signal-background interference
in $t\bar t$ at LO, leading to a suppression of $ -420$ fb while the
$hh$ signal with mass $m_{H_3}=694$~GeV is enhanced by a factor 2.3 due to
signal-signal interference. 
Also the for this benchmark point phenomenologically
  more relevant resonant production of $H_2$ with a mass of 691 GeV is
  slightly enhanced due to signal-signal interference.
Analogous to our results for C2HDM type 1,
this indicates that usual  SM search channels might have suppressed
sensitivity and new physics could indeed be observed in di-Higgs
production first. Although the $hh$ cross section is significantly
smaller compared to the $t\bar t$ rate, the relative enhancement
through signal-signal interference magnifies the importance of
di-Higgs searches at the LHC. Another, milder example along this line
is given by {\texttt{BP3}} (Fig.~\ref{fig:bp10544}) where  the $t\bar t$
resonance structure is significantly distorted to a peak-dip
combination,  while $pp\to hh$ remains as a Breit-Wigner peak at a
slight 30\% suppression (LO-)rate.   
The masses of $H_2$ and $H_3$ in Figs. ~\ref{fig:bp17783} and ~\ref{fig:bp10544}
are quite similar, and indeed the curves on the left-hand side of the figures ($t {\overline {t}}$) are quite
similar for the two masses.  It is interesting, however, the difference between the curves for these two mass
points on the right-hand side of the figures ($hh$) is clearly visible, due to the difference in the signal-signal
interference that can be observed in $pp\to hh$ production.

The qualitative behaviour that we have found so far is related to the
fact that the scalar spectra are quite compressed  in the examples we
have considered so far, where we have
\begin{equation}
\label{eq:compre}
|m_{H_2}-m_{H_3}| < 0.1\,m_{H_3}\,.
\end{equation}
Given the finite experimental resolution, states that are nearly degenerate are difficult to resolve. Nonetheless, if an excess
is observed in the future, signal-signal interference is an important effect that
needs to be taken into account if the excess is interpreted along the lines of two nearly degenerate states in a C2HDM.

It is also interesting to discuss the opposite situation
when the two scalars lie further apart in mass. Under these
circumstances, signal-signal interference is largely absent in $hh$
production, while signal-background interference again distorts or
even removes the resonance structure in $H_i\to t\bar t$. When
$|m_{H_2}-m_{H_3}| > 0.1\,m_{H_3}$ we can therefore conclude that
$\sigma$-times-branching ratio estimates provide an accurate
description of the $hh$ phenomenology while the
sensitivity in $t\bar t$ is 
again overestimated, as illustrated in Fig.~\ref{fig:bp1019}. For this
point {\texttt{BP4}} (see Tab.~\ref{tab:bench}),  we again observe a
large signal -background interference leading to $\simeq -300$ fb for
the on-shell region according to Eq.~\eqref{eq:intrange} in $t\bar t$
with $pp\to hh$ production remaining stable

\subsubsection{The CxSM}
We finally turn to the relevance of CP violation in the above
context. To this end, it is helpful to consider the CxSM of
Sec.~\ref{sec:cxsm}. Again, a wide range of mass spectra can be
obtained that are consistent with the scan outlined in
Sec.~\ref{sec:cxsmscan}. This includes compressed spectra according to
Eq.~\eqref{eq:compre} through selecting appropriate parameters of the
extended Higgs potential that pass LHC constraints.

As we are dealing with a model that anti-correlates SM Higgs coupling
consistency with the size of the exotic scalar cross sections, $H_i\to
t \bar t $ production is typically not phenomenologically
relevant. This is also the reason why interference effects can be very
large in this case as the signal cross section can be
negligible. Nonetheless it is interesting to observe that the di-Higgs
cross section is not influenced dramatically and qualitatively
identical when we compare degenerate and non-degenerate mass spectra
in Fig.~\ref{fig:singlet}. All points are around $m_{H_i}\simeq
400~\text{GeV}$ where the SM di-Higgs cross section reaches a
maximum. The second non-SM-like Higgs boson $H_j$ has
  either a mass close to $m_{H_i}$ according to our criteria
  Eq.~(\ref{eq:compre}) (left plot) or is far apart (right plot).
In this scenario we deal indeed  with only mild
signal-background and signal-signal interference, {\it{i.e.}}
$\sigma$-times-branching ratio expectations provide an accurate estimate
of the $hh$ phenomenology in this scenario. This can be contrasted with the
C2HDM phenomenology: When we switch on CP violation, we typically
create the possibility of gauge-phobic Higgs bosons
(we remind the reader that in the CxSM we only deal
  with neutral CP-even Higgs bosons), which then
preferably couple to top quarks when the gluon fusion production mode
is large. Additionally, when the trilinear couplings are large enough,
$H_i\to hh$ can become sizeable while $H_i\to t\bar t$ is opening up
to large signal-background interference effects.

\section{Conclusions}
\label{sec:conc}
Signal-background interference is a phenomenologically important feature in resonance searches in top final states. In this paper we have shown that  the signal-background interference that the ATLAS and CMS experiments are already considering in setting limits on  models of the kind discussed in this work needs to be extended to signal-signal interference effects, in particular when considering di-Higgs production as a BSM discovery tool. 

While constructive enhancements in the $t\bar t$ final states with
large cross sections could be a smoking gun of BSM physics in the near
future~(see the recent \cite{Djouadi:2019cbm}), an equally interesting
outcome is the decrease of the $t\bar t $ signal due to
signal-background interference that is correlated with an enhancement
of the resonant di-Higgs production rate as a consequence of
significant signal-signal interference. The latter becomes important
when extra scalar resonances are heavy and reasonably close in
mass. This is a region of the C2HDM parameter space which naturally
arises given current LHC observations. Our findings therefore indicate
that the relevance of di-Higgs final states in the context of the
CP-violating C2HDM has so far been underestimated. We provide a number
of benchmark scenarios that highlight the phenomenological relation of
$t\bar t $ and $hh$ resonance searches. We have limit ourselves to stable top and Higgs boson
final states in this work. $t\bar t$ resonance searches are well underway (e.g.~\cite{Sirunyan:2019wph}), and
the large SM backgrounds that can be limiting factors of di-Higgs searches have been shown to be manageable, see
the recent CMS analysis of Ref.~\cite{Sirunyan:2018two}.

In actual experimental
analyses, the discrimination of nearly degenerate resonances is hampered 
by the finite resolution that can be obtained in $t\bar t$ or di-Higgs resonance
searches. Yet, interference
effects are important to reach to correct microscopic parameter interpretation.

It is worthwhile noting that similar effects are not present in
simpler scenarios such as, {\it{e.g.}}, a complex scalar portal
extension of the Higgs sector. In the mass region where the effects
discussed in this paper work efficiently, the cross sections in $t\bar
t $ are too suppressed to be phenomenologically relevant, which de
facto de-correlates the $t\bar t$ and $hh$ channels. In the CxSM the 
$t\bar t$ channels are suppressed (see also \cite{Muhlleitner:2017dkd}) while
in the C2HM inteference effects distort the shapes of the resonances. While the latter
makes discoveries in the $t\bar t$ channel difficult the discovery of
a new scalar resonance with a significant cross section in $t\bar t$
would strongly discriminate between the C2HDM 
and the CxSM.

\noindent {\bf{Acknowledgements}} ---
P.B. acknowledges financial support by the Graduiertenkolleg “GRK
1694: Elementarteilchenphysik bei h\"ochster Energie und h\"ochster
Pr\"azision. M.M. acknowledges support by the Deutsche
Forschungsgemeinschaft (DFG, German Research Foundation) under grant
396021762 - TRR 257. 
S.D. is  supported by the U.S. Department of Energy under Grant
Contract de-sc0012704.  
C.E. is supported by the UK Science and Technology Facilities Council
(STFC) under grant ST/P000746/1. 
This work was supported by the Munich Institute
for Astro- and Particle Physics (MIAPP) of the DFG Excellence Cluster
Origins (\url{www.origins-cluster.de}). 

\enlargethispage{-5\baselineskip}
\bibliography{paper.bbl}

\end{document}